# Crystal Growth and characterization of possible New Magnetic Topological Insulators $FeBi_2Te_4$

Ankush Saxena[1,2], Poonam Rani[1], S. Patnaik[3], Vipin Nagpal[1], I. Felner[4] and V.P.S. Awana[1,2,*]

[1] *National Physical Laboratory (CSIR), Dr. K. S. Krishnan Road, New Delhi-110012, India*

[2] *Academy of Scientific and Innovative Research (AcSIR), Ghaziabad-201002, India*

[3] *School of Physical Sciences, Jawaharlal Nehru University, New Delhi-110067, India*

[4] *Racah Institute of Physics, The Hebrew University, Jerusalem, 91904, Israel*

**Abstract**

Here we report successful single crystal growth of new possible magnetic topological insulator (MTI) $FeBi_2Te_4$ by self-flux method via vacuum encapsulation process. The detailed Rietveld analysis of Powder XRD data shows the as grown MTI crystal to be mainly dominated by $FeBi_2Te_4$ phase along with minority phases of $Bi_2Te_3$ and FeTe. Scanning electron microscope (SEM) image shows the morphology of as grown MTI single crystal to be of layered type laminar structure. Raman spectroscopy of the crystal exhibited three distinct phonon modes at 65, 110, and $132 cm^{-1}$ along with two split secondary modes at 90, and $144 cm^{-1}$. The secondary split modes are result of FeTe intercalation in $Bi_2Te_3$ unit cell. Magneto-resistance (MR%) measurement has been performed at different temperatures i.e. 200K, 20K and 2K in applied magnetic fields up to ±12 Tesla, which showed very low MR in comparison to pure $Bi_2Te_3$ crystal. Temperature dependence of DC magnetization measurements show the $FeBi_2Te_4$ crystal to be mainly of ferromagnetic (FM) or ferri-magnetic nature above 295 K, albeit a secondary weak magnetic transition is seen at 54-46K as well. Detailed isothermal magnetization (MH) results showed that FM saturation moment at 295K is 0.00213emu/g, which is nearly invariant till 400 K. Summary, we had grown an MTI $FeBi_2Te_4$ single crystal, which may be a possible entrant for Quantum Anomalous Hall (QAH) effect at room temperature or above.

*Corresponding Author
Dr. V. P. S. Awana: E-mail: awana@nplindia.org
Ph. +91-11-45609357, Fax-+91-11-45609310
Homepage: awanavps.webs.com

**Introduction**

Topological Insulators (TI) are among the newest wonder materials with various novel physical properties, including highly conducting surface/edge electronic states along with the insulating bulk [1, 2]. These highly mobile conducting states are Spin Orbit Coupled (SOC) as well protected by Time Reversal Symmetry (TRS) [3, 4]. The most popular bulk TIs are $Bi_2Se_3$, $Bi_2Te_3$ and $Sb_2Te_3$. As far as their structure is concerned, these exhibits a layered structure composed of quintuples, containing five atomic layers arranged in Te/Se-Bi/Sb-Te/Se-Bi/Sb-Te/Se fashion. These are further bonded by ionic-covalent bonds. Each quintuple layer is a reverse image of its adjacent quintuple layer and separated by weak van der Waals forces [2, 3]. The presence of van der Waals gap in between the quintuple layers makes the compound easily cleavable and susceptible to various intercalations [2, 3, 5]. Suitable intercalation of some doping materials into these van-der Waals gap results in various quantum phenomenon including superconductivity [6-10] and Quantum Anomalous Hall (QAH) effect [11, 12].

The reports for QAH effect are only scant, and that also for a particular Magnetic Topological Insulator (MTI) i.e., $MnBi_2Te_4$ [11-14]. For observation of QAH effect, one needs to insert a magnetically ordered layer between van der Waals gaps of bulk TI, viz. $MnBi_2Te_4$. Here an anti-ferro-magnetically (AFM) ordered ($T_N = 20K$) MnTe layer is inserted between van der Waals gaps of bulk TI $Bi_2Te_3$ [11-14].

The quest is for elevated temperature and low magnetic field MTI. That means the inserted magnetic layer must be near room temperature magnetically ordered. Besides, $MnBi_2Te_4$, one such example could be $FeBi_2Te_4$. Here $FeBi_2Te_4$ represents FeTe + $Bi_2Te_3$. The former is AFM ordered ($T_N = 80K$) FeTe [15, 16] and later is bulk TI $Bi_2Te_3$. We managed successfully to grow single crystal of $FeBi_2Te_4$ and to best of our knowledge no report on crystal growth or characterization of $FeBi_2Te_4$ exists in literature. On the other hand we failed to grow other possible TMI as well viz. $CoBi_2Te_4$ and $NiBi_2Te_4$. The resultant material did not even look crystal alike as usual silvery shining as shown in Fig. 1a. In case of $FeBi_2Te_4$, we got a silvery shinning crystal with reasonably good X-ray diffraction (XRD) similar to that of $MnBi_2Te_4$ [11-14]. Interestingly, the compound seems to become ferromagnetic (FM) or ferry-magnetic above 250 K. These are preliminary results and more need to be done in terms of detailed characterization of the obtained crystal by Transmission Electron Microscope to assure its crystalline nature along with detailed Hall measurements. In this letter we report the possible crystallisation of new MTI, $FeBi_2Te_4$ along with its basic transport and magnetic properties. We believe this study will encourage the condensed matter physicist and solid-state chemist communities to follow up and establish a near room temperature (250K) FeBi$_2$Te$_4$ Quantum Anomalous Hall (QAH) effect.

**Experimental**

The constituent elements Fe, Bi and T$_c$ (at least of 3N purity) were weighed (1gram) in stoichiometric ratio and grinded in a glove box filled with Argon. After grinding, the powder was converted into pallets, sealed in a tube under the pressure of $10^{-5}$mb and inserted in a controlled furnace at high temperature. First, the temperature was raised to 950˚C with a heating rate of 120$^0$C/hour, which is hold for 12 hours. Then, the temperature was decreased slowly at the rate of 1˚C/hour down to 600˚C. This temperature was kept for 12 hour followed by normal cooling to room temperature. A schematic details of the heat treatment are given in Fig. 1(a) and the photograph of as grown crystal (MTI) is shown in Fig. 1(b).

Powder X-ray diffraction pattern (PXRD) of the gently crushed part of crushed crystal was taken on Rigaku X-ray diffractometer in the range 10˚ to 80˚ of 2θ˚ at scan rate of 2˚/min. Bruker make Scanning Electron Microscope (SEM) was used to visualize the morphology of the studied crystal. Raman spectrum of the crystal piece is taken on LabRam HR800-JY equipped with a laser source of 514 nm. The spectrum was taken in wave number range of 50cm$^{-1}$ to 400cm$^{-1}$. Resistivity versus temperature measurements were done on a Quantum Design (QD) Physical Property Measurement System (PPMS) in temperature range of 300K down to 5K, and the DC magnetization studies were performed in QD MPMS magnetometer. The DC measurements under external field (H) of 1 kOe, at the temperature 5-300 K and 250-485 K intervals were done on two separate pieces cut from the same crystal.

**Results and Discussion**

Fig. 1(c) shows the typical SEM picture of the as grown FeBi$_2$Te$_4$ crystal. The layered structure is clearly visible, which is known and reported earlier for other bulk topological insulator single crystals [1-4, 17]. Fig. 2 depicts the PXRD of the gently crushed FeBi$_2$Te$_4$ crystal. Three options including their Rietveld analysis are depicted. (a) The lower panel shows the PXRD of FeBi$_2$Te$_4$ crystal with solely, (b) the middle the co-existing phases of FeTe and FeBi$_2$Te$_4$ and (c) the upper panel with FeTe and Bi$_2$Te$_3$ only (without FeBi$_2$Te$_4$). These three fitting panels are required, because the single phase FeBi$_2$Te$_4$ option (lower panel) could not account for a low angle peak at around 2Θ = 18° and for some other small intensity peaks. The goodness of fitting parameters ($\chi2$ =11%) with sole FeBi$_2$Te$_4$ phase is poor. This interpretation is quite similar to that of MnBi$_2$Te$_4$ [11-14]. However, for option b when both FeBi$_2$Te$_4$ and FeTe are considered, the fitting parameter $\chi^2$ is reduced to around 7.15%. The lattice parameters thus calculated for the main FeBi$_2$Te$_4$ (70%) phase are $a$ = 4.3926(4)A° and $c$ = 42.6944(3)A° and for the cubic FeTe (30%) $a$ = 3.826(1)A°, this agrees with earlier reports [15, 16]. This option is supported by our magnetic and Raman studies reported below. Further, when the PXRD Rietveld analysis is done with FeTe and Bi$_2$Te$_3$ phases (upper panel), $\chi^2$ is reduced to around 5.25%. Then, the lattice parameters calculated for the majority (75%) phase Bi$_2$Te$_3$ are $a$ = 4.3851(4)A° and $c$ = 30.4952(4)A° and for the minority FeTe (25%) $a$ = 3.826(1)A° as in option b. Details of various phases are given in Table 1. As stated above, our preferred option (b) is that the studied FeBi$_2$Te$_4$ crystal contains also 30% of FeTe an extra phase albeit pure single crystal of FeBi$_2$Te$_4$ being devoid of FeTe and Bi$_2$Te$_3$ phases, is yet warranted.

Fig. 3 shows the Raman spectrum of as grown FeBi$_2$Te$_4$ crystal. Three distinct peaks are seen at 65, 110, and 132 cm$^{-1}$ (assigned as 1, 3 and 4) Further, two secondary split modes are seen at 90, and 144 cm$^{-1}$ (assigned as 2 and 5) The general appearance of Raman spectrum in Fig. 3 is similar to that as reported earlier for MnBi$_2$Te$_4$ [11, 14]. The main phonon modes at 65, 110, and 132 cm$^{-1}$ are known to be due to three different Raman vibrational modes namely A$^1$g$_1$, A$^1$g$_2$, and Eg$^2$ for Bi$_2$Te$_3$, and are comparable to the earlier reported result [17, 18]. In case of MnBi$_2$Te$_4$, various split modes are seen in ref. 12 and are attributed to the van-der Waals gap inserted MnTe bonds with parent Bi$_2$Te$_3$ in MnBi$_2$Te$_4$ unit cell. Further, it is suggested that not only the MnBi$_2$Te$_4$ but various other possible crystallographic arrangements depending upon super structures are possible, viz. MnTe + nBi2Te3, with n = 1, 2, 3, as MnBi$_2$Te$_4$, MnBi$_4$Te$_7$, and MnBi$_6$Te$_{10}$ phases [12]. A more consolidate and detailed over view of the Raman data analysis for similar van-der Waals gap inserted TI i.e. PbB$_2$Te$_4$ (PbTe+Bi$_2$Te$_3$) are reported recently, highlighting the splitting of parent phonon modes due to intercalation [19]. Keeping in view the reported Raman spectroscopy results on MnBi$_2$Te$_4$ [12, 14], PbB$_2$Te$_4$ [19], and the observation of split secondary modes for FeBi$_2$Te$_4$ crystal, we may safely conclude that to a large extent FeTe is inserted in van-der Waals gaps of parent Bi$_2$Te$_3$ layers in the studied FeBi$_2$Te$_4$ crystal.

Fig. 4 shows the isothermal magneto resistance (RH) of FeBi$_2$Te$_4$ up to H=12 Tesla at 200, 20 and 2 K. At 200 K the studied FeBi$_2$Te$_4$ crystal shows positive non saturating magneto resistance (MR%) of up to 40% under 12 Tesla. Here MR% is calculated as MR% = [(R$_H$-R$_0$)/(R$_0$)]x100. Albeit quite small but an interesting feature is seen in low fields of below say 0.5 Tesla, where MR is negative in both directions. This feature is very similar to that as observed below the AFM ordering temperature (25 K) of Mn spins in MnB$_2$Te$_4$ [11-14]. It is known that MnBi$_2$Te$_4$ exhibits Quantum Anomalous Hall (QAH) effect below 25 K [11-13]. Here, at 200 K a similar phenomenon, i.e. signature of Quantum Anomalous Hall (QAH) effect albeit small, is seen. Surprisingly at 20 and 2 K, the positive MR magnitude increases to around 5% and 2% only at 12 Tesla, and the low field negative MR feature is also missing. It is expected that deep below the magnetic ordering of van-der Waals gap due to inserted magnetic FeTe layers Bi$_2$Te$_3$, MTI would demonstrate better QAH effect. This does not look to be the case. Maybe that QAH is more expressed close to magnetic transition to a FM state as observed hereafter. May it be the QAH effect be seen near FM ordering i.e. exactly near room temperature. However, yet a lot more needs to be done in terms of phase purity and long-range magnetic ordering of the new MTI, FeBi$_2$Te$_4$.

Figs 5 (a,b) show the ZFC and FC plots of FeBi$_2$Te$_4$ measured at 106 Oe and 1 kOe (0.1 T) respectively. All plots show a moderate increase at low temperatures, indicating the presence of tiny amount of a paramagnetic extra phase. In addition, three anomalies are observed: (i) A pronounced peak in the ZFC branches at 54 and 46 K respectively. (ii) A tiny rise at 118K and (iii) both ZFC and FC plots increase sharply around 250 K (T$^{mag}$) and tend to emerge at 295 K. On a second piece of the same crystal, an extended ZFC (at H=1kOe) measurement up to 385 K was performed as depicted in Fig. 5b, inset. (i) Due to the relatively large magnitude of ZFC peak at 54K Fig. 5a, the FC branch crosses the ZFC one, thus at certain temperature range ZFC>FC. This is a unique phenomenon seldom observed [20,21]. On the other hand, at 1 kOe, the peak magnitude at 48 K is lower and the normal FC>ZFC behaviour up to RT exists. Note also the bumps in both FC branched around these temperatures. The reproducibility of peak in the second ZFC plot (at 1kOe), excludes the peculiar observations of ZFC>FC observed e.g. in amorphous carbon in the past, in which this phenomenon was observed in the first ZFC run only, irreproducible and washed up in the second ZFC run [20,21].

The origin of the peaks around T ~ 50 K is debatable. The peaks may originate from adsorbed traces of solidified oxygen [22]. Actually, traces of solidified oxygen are visible, but they are negligible in single crystals in which the bulk properties are dominated [20]. It is important to note that the presence of oxygen always appears as sharp peaks at the same temperature in both ZFC and FC branches [20,21]. Alternatively, these peaks may arise from the presence of the AFM FeTe (T$_N$=60 K) extra phase as suggested above (option b). The lower temperatures obtained, are probably caused by the possibility that the FeTe extra- phase is not stoichiometric, or alternatively that this phase is doped by tiny amount of Bi. Anyhow, this AFM transition is extremely sensitive to the external field and shifts by 8 degrees (54 and 46 K) when H increases from H=106 to 1 kOe.

(ii) The deviation at 118 K is certainly caused by the presence of tiny amount of magnetite ($Fe_3O_4$) the so-called the Verway transition. The rise in the ZFC branch at 118 K and 106 Oe (Fig. 5a) is $7.3*10^{-8}$emu/g Oe. For sake of comparison, we measured pure bulk $Fe_3O_4$ under similar conditions, for which the rise at 118K accounts to $7.8*10^{-2}$emu/g Oe. That means that the amount of $Fe_3O_4$ in the sample is around 1ppm.

(iii) Under all measured fields, a sharp increase of the magnetization exists at around 250 K. Fig. 5b (inset) shows that the magnetization (measured on a second piece) is almost constant in 300-385 K intervals. Thus, we tend to believe that in the major part of $FeBi_2Te_4$, a magnetic phase transition from AFM (for T< 250 K) to ferro-magnetic or ferrimagnetic occurs above 250 K, and that its magnetic transition temperature is well above 400 K. This observation is consistent with the bifurcation of ZFC(T) and FC(T) branches around RT and also supported by the isothermal magnetization M(H) exhibited in Figs. 6-7.

It is readily observed, that the M(H) curves first increase up to 1–1.5kOe, tends to saturate and then linearly decrease up to 50 kOe. The experimental M(H) plot clearly reveals an admixture of magnetic and diamagnetic components and can be fitted as: $M(H)_{exp} = M_{Sat} +(- \chi H)$, where the saturation moment $M_{Sat}$ is the intrinsic magnetic phase contribution, and $-\chi H$ is the linear diamagnetic contribution stem e.g. from the sample holder. Fig. 7 demonstrate this procedure for T=295 K. The $M_{Sat}$ values obtained are: 0.035, 0.0011, 0.00098 and 0.00213emu/g for T= 5,100, 200 and 295 K respectively. Note that $M_{Sat}$ for 100 and 200 K are almost the same and twice as much at 295 K. Due to these relatively low $M_{Sat}$ vales obtained, the diamagnetic signals are easily pronounced. That indicates that below 250 K the major phase of $FeBi_2Te_4$ is AFM ordered. All M(H) plots were also measured under negative external fields from which the various coercive fields ($H_C$) shown in Fig. 6 (inset) can be deduced. Similar to the $M_{Sat}$ behaviour, $H_C$ =390(10) Oe at 5K, decreases to 170(10) Oe at both 100 and 200K and that increases (by a factor of 7) to 1150(10) Oe at 295K. That confirms our statement that around 250K (Fig. 5) a magnetic phase transition from AFM to FM (or ferrimagnetic) occurs.

In conclusion, we have grown an MTI $FeBi_2Te_4$ single crystal, having a ferromagnetic magnetic ordering well above 400 K, which may be a possible new entrant for Quantum Anomalous Hall effect at elevated temperatures.

Table 1: Rietveld refined structural parameters including, Lattice constants, space group, co-ordinates and phase proportions for various phases of FeBi$_2$Te$_4$.

|  | Bi$_2$Te$_3$+FeTe | | FeBi$_2$Te$_4$+FeTe | | FeBi$_2$Te$_4$ |
|---|---|---|---|---|---|
|  | Bi$_2$Te$_3$ | FeTe | FeBi$_2$Te$_4$ | FeTe | FeBi$_2$Te$_4$ |
| **Cell (a=b)** | 4.3851(4) | 3.826(1) | 4.3922(1) | 3.826(1) | 4.3926(4) |
| **c** | 30.4952(4) | 3.826(1) | 42.6944(3) | 3.826(1) | 42.6944(3) |
| **Angle (α=β)** | 90 | 90 | 90 | 90 | 90 |
| **γ** | 120 | 90 | 120 | 90 | 120 |
| **Space Group** | R -3 m | P 4/n m m | R -3 m | P 4/n m m | R -3 m |
| **Percentage** | 49.77 | 50 | 70 | 30 | 100 |

**Figure Captions:**

Fig. 1 (a) Schematic of heat treatment schedule, (b) the photograph and (c) SEM image of as grown $FeBi_2Te_4$ crystal

Fig. 2 Powder x-ray diffraction (PXRD) of the gently crushed $FeBi_2Te_4$ crystal, lower panel is the Reitveld analysis of PXRD with in sole $FeBi_2Te_4$ phase and the upper one with two co-existing phases of FeTe and $Bi_2Te_3$.

Fig. 3 Raman spectrum of as grown $FeBi_2Te_4$ crystal

Fig. 4 Isothermal magneto resistance (RH) at 2, 20 and 200K in applied field of up to 12 Tesla of as grown $FeBi_2Te_4$ crystal

Fig. 5 a, Magnetic moment versus temperature (MT) at 106 Oe in both ZFC and FC modes from 300K down to 5K, Note the sharp peak at 54 K and that around the peak ZFC>FC. b. Magnetic moment versus temperature (MT) at 1000 Oe in both ZFC and FC modes from 300K down to 5K, the inset shows the same at high temperatures of up to 390K in ZFC mode for the as grown $FeBi_2Te_4$

Fig. 6: Isothermal magnetization (MH) plots at 5, 100, 200 and 295K for up to 50K Oe field for as grown $FeBi_2Te_4$. The inset shows the temperature dependence of coercive field values.

Fig 7 Isothermal magnetization (MH) plot at 5 K along with –ve susceptibility of the parent $Bi_2Te_3$ and thus the resultant for as grown $FeBi_2Te_4$

Fig. 1 (a)

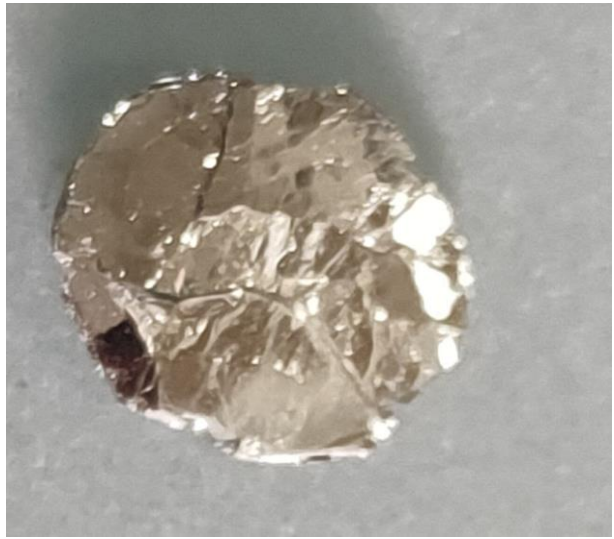

Fig. 1 (b)

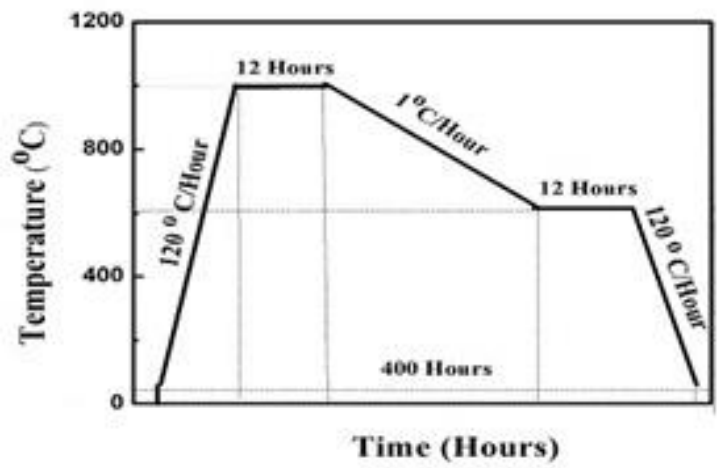

Fig. 1 (c)

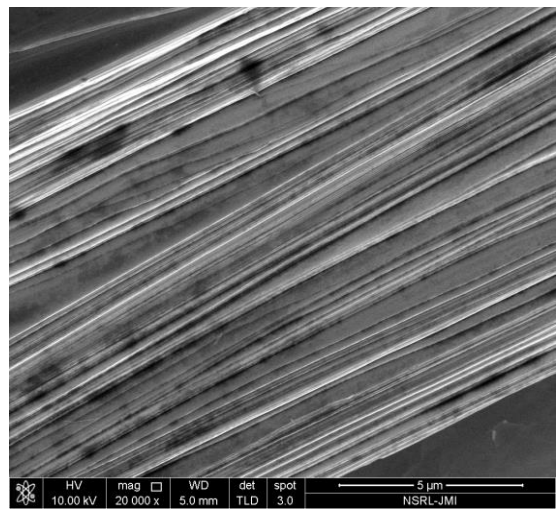

Fig. 2

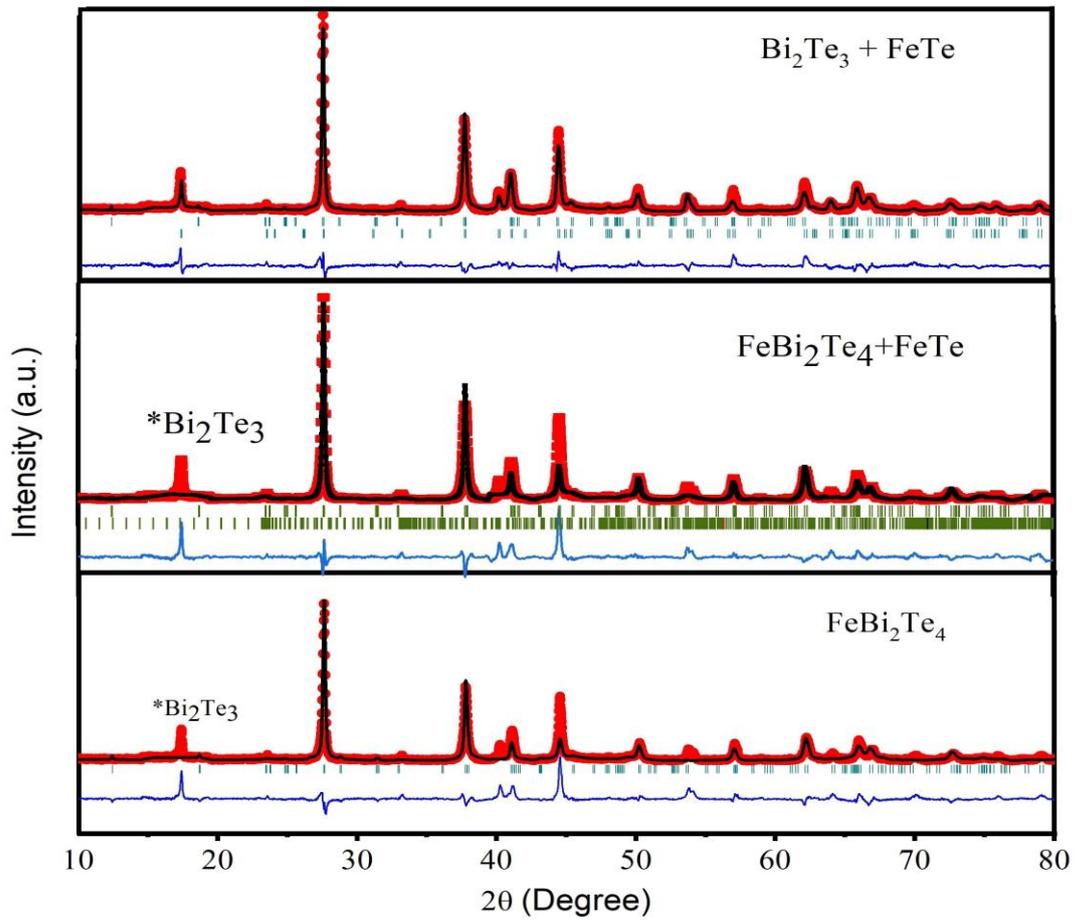

Fig. 3

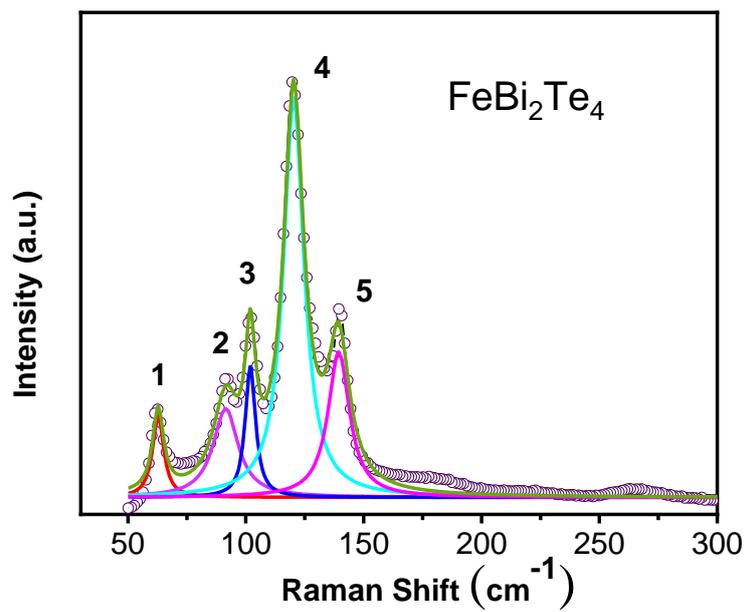

Fig. 4

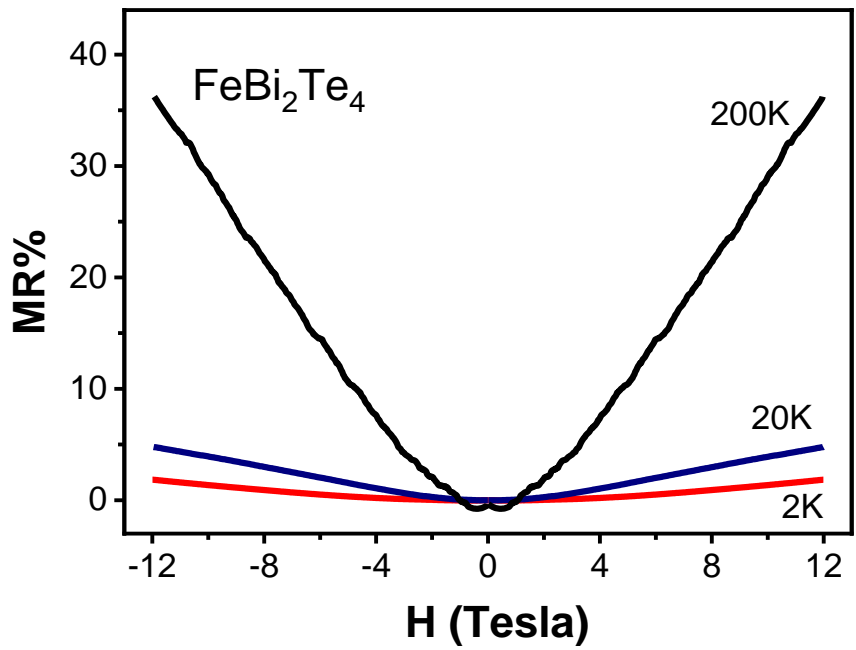

Fig 5:

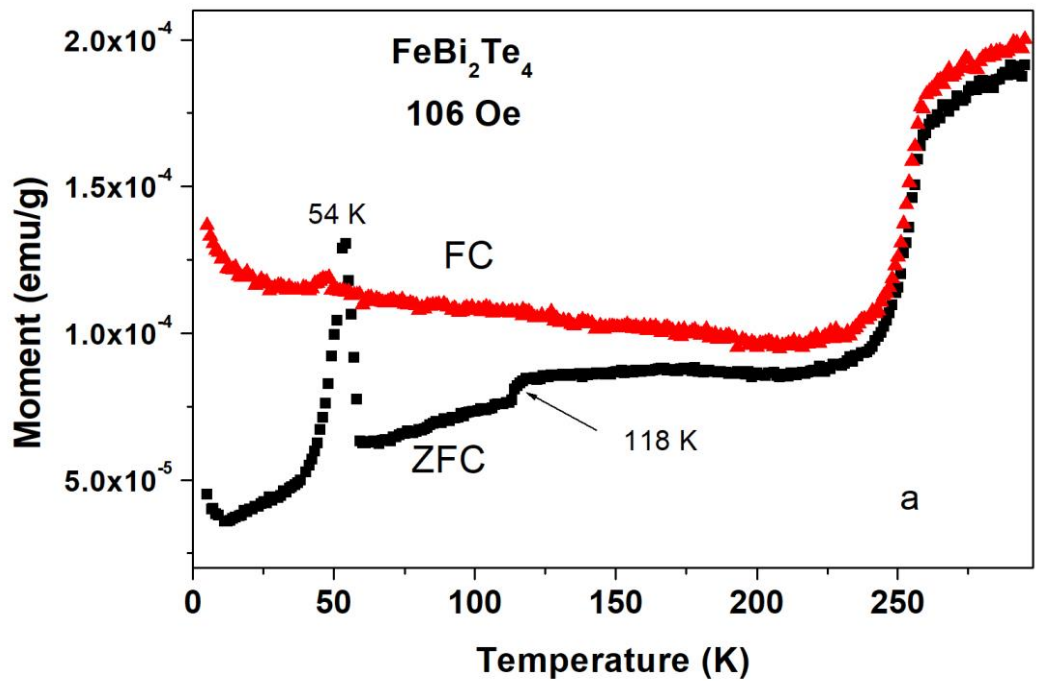

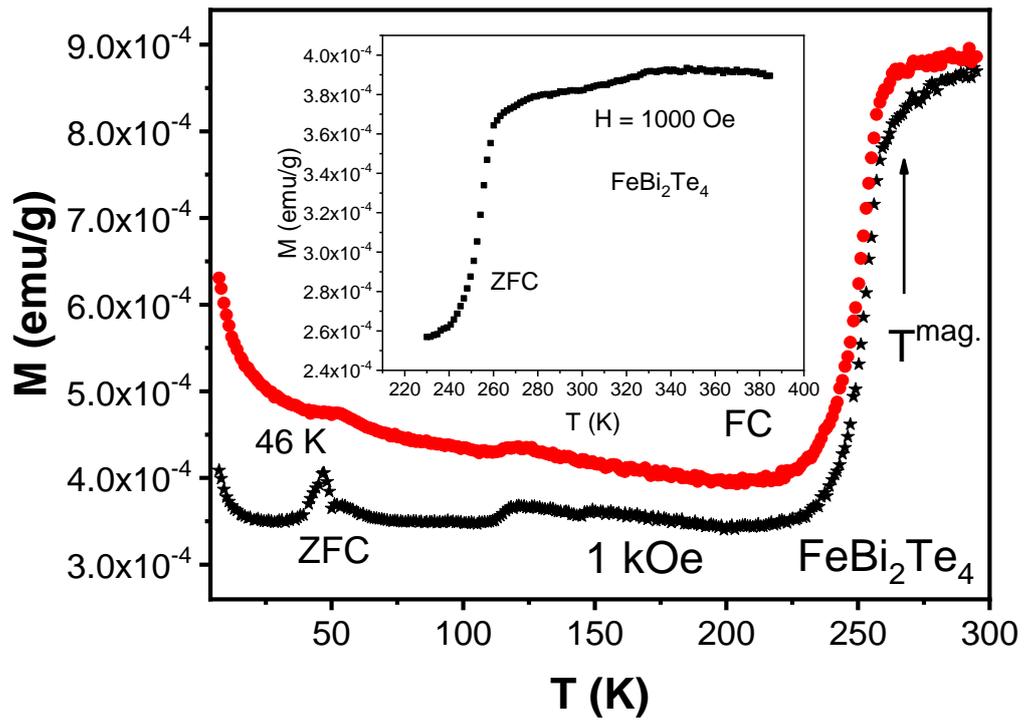

Fig 6:

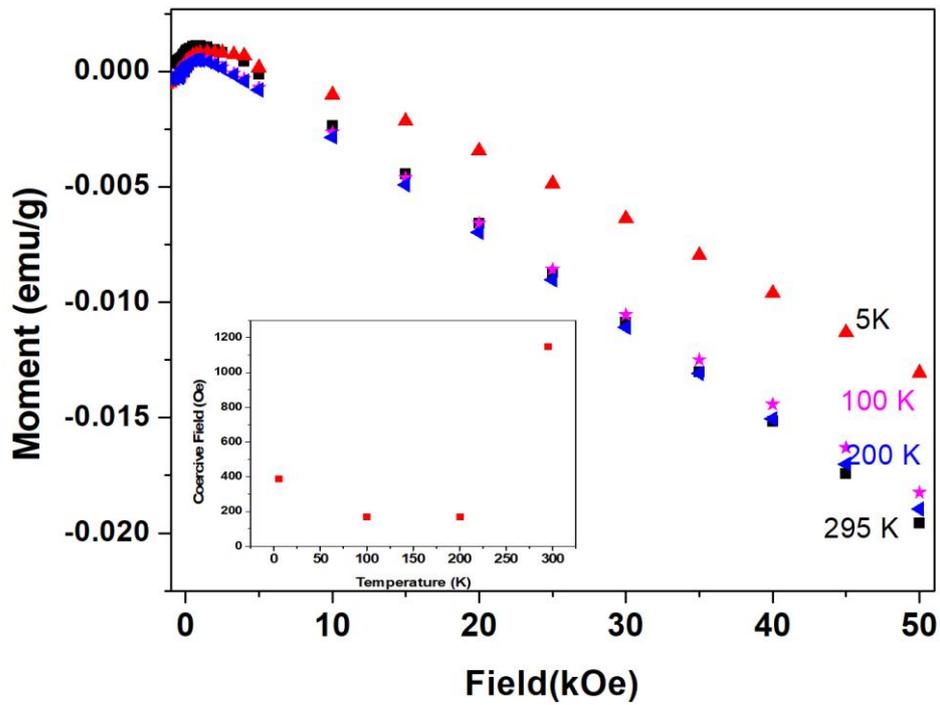

Fig. 7

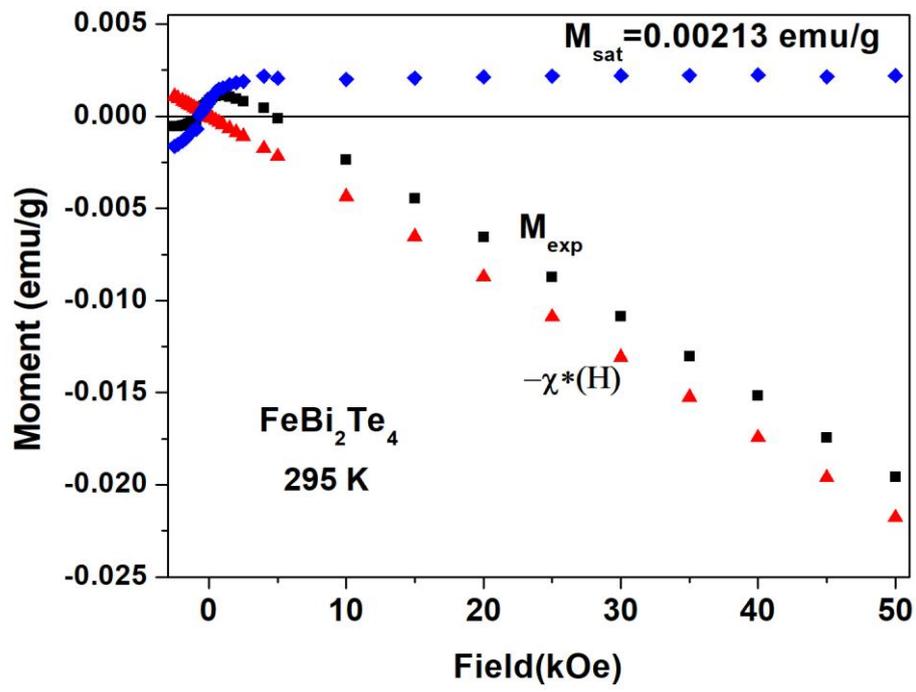